\documentclass[5p,twocolumn,times,number]{elsarticle}

\usepackage{amssymb}
\usepackage{boxedminipage}
\usepackage{upgreek}

\usepackage{lineno}

\journal{Nuclear Instruments and Methods A}

\begin{document}

\begin{frontmatter}



\title{Effect and suppression of parasitic surface damage in neutron irradiated CMOS Monolithic Active Pixel Sensors}


\author[LabelUniFFM]{M. Deveaux}
\author[LabelUniFFM]{D. Doering}
\author[LabelGSI]{P. Scharrer}
\author[LabelUniFFM,LabelGSI]{J. Stroth}

\address[LabelUniFFM]{Goethe-University Frankfurt, Max-von-Laue-Str.1, D-60439 Frankfurt/M}
\address[LabelGSI]{GSI, Plankstr.1 , D-64291 Darmstadt}

\begin{abstract}
CMOS Monolithic Active Pixel Sensors (MAPS) were chosen as sensor technology for the vertex detectors of STAR, CBM
and the upgraded ALICE-ITS. They also constitute a valuable option for tracking devices at future e+e- colliders. Those applications require a substantial tolerance to both, ionizing and non-ionizing radiation. To allow for a focused optimization of the radiation tolerance, prototypes are tested by irradiating the devices either with purely ionizing radiation (e.g. soft X-rays) or the most pure sources of non-ionizing radiation available (e.g. reactor neutrons). In the second case, it is typically assumed that the impact of the parasitic $\gamma$-rays found in the neutron beams is negligible. We checked this assumption by irradiating MAPS with $\gamma$-rays and comparing the radiation damage generated with the one in neutron irradiated sensors. We conclude that the parasitic radiation doses may cause non-negligible radiation damage. Based on the results we propose a procedure to recognize and to suppress the effect of the related parasitic ionizing radiation damage.  

\end{abstract}


\end{frontmatter}


\section{Introduction}
CMOS Monolithic Active Pixel Sensors (MAPS) have demonstrated excellent performances for charged particle tracking. Devices with a surface of up to $2~\rm cm^2$ reached routinely detection efficiencies of $>99.9\%$, a spatial resolution of better than $5~\rm \upmu m$ and likely a tolerance to neutron fluencies of up to $3 \cdot 10^{14}~\rm n_{eq}/cm^2$ \cite{Paper:MIMOSA-26,Paper:1e14}. They were therefore chosen for the vertex detectors of STAR and CBM
and are foreseen to equip the upgraded ALICE-ITS. They also constitute a valuable option for tracking devices at future e+e- colliders. Improving their radiation tolerance to the needs of those experiments is subject of a common R\&D program of the PICSEL group of IPHC Strasbourg and the IKF Frankfurt. 

To study their radiation tolerance, the MAPS are irradiated either with  X-rays or reactor neutrons and tested hereafter. The sources are chosen knowing that the energy of soft X-rays is sufficient to ionize Si-atoms but insufficient to displace them. The uncharged $\sim 1~\rm MeV$ reactor neutrons rarely ionize the atoms but displace them from their position in the crystal lattice. Neutrons create therefore almost exclusively displacement damage in the active silicon volume of the pixel (bulk damage) while X-rays break irregular bonds at the surfaces between the $\rm Si$ and the $\rm SiO_2$-structures, which are widely found as isolator in CMOS devices (surface damage).

\begin{figure}[t]
\begin{center}   
   
    \includegraphics[viewport= 4cm 6cm 22cm 19cm, clip, width=.9\columnwidth]{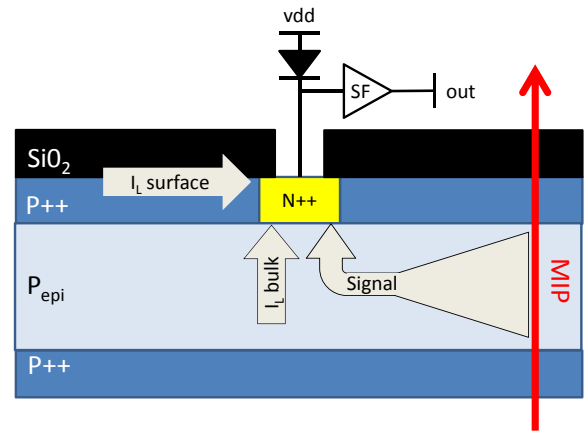}
   \caption{\label{Figure:Geometry} Anticipated locations of the sources of leakage currents caused by bulk damage ($\rm I_{L}~bulk$ and surface damage ($\rm I_L~surface$) in a SB-pixel. SF= Source Follower, MIP = Minimum Ionizing Particle.}
	\end{center}
\end{figure}

Figure \ref{Figure:Geometry} displays the operation principle of a MAPS with a so-called self-bias (SB-)pixel and the consequences of radiation damage on this pixel. In the P-doped active volume (P$\rm_{epi}$) of the pixel, impinging particles excite electrons from the valence band to the conduction band of the silicon. Those electrons diffuse in the active volume until they are collected by a diode formed from the P$_{epi}$ and an N-Well implantation. Here, they discharge the parasitic capacity of the diode. The related voltage drop is sent via a source follower to an external ADC/discriminator. Hereafter, the signal is cleared slowly by means of a high-ohmic forward biased diode\footnote{The properties of the self-bias pixel, its amplifier and related radiation effects are described in more detail in \cite{Paper:SB-pixel}.}. 

Bulk damage reduces the life time of the minority charge carriers and thus increases the probability that the electrons recombine before being collected.
 Moreover, it eases the thermal generation of minority charge carries and increases the dark current and thus the shot noise of the collection diode.  A similar increase is caused by surface damage. In the latter case, the generation centers are likely formed by broken bonds located at the $\rm Si/SiO_2$ interface nearby the collection diode. 
 
Finally, ionizing radiation creates meta-stable positive charge carriers located at this interface. The related electric fields may deform the bands of the silicon, open conduction channels and modify the characteristics of field effect transistors.

The leakage currents caused by both, ionizing and non-ionizing radiation, may be substantially reduced by means of cooling. Moreover, reducing the integration time may alleviate the related shot noise. However, the reach of both measures is finite as room temperature operation is desirable for many applications and as the integration time of the sensors cannot be reduced below a certain limit. As a complementary measure, one might consider to reduce the leakage currents by means of improved sensor design. This goal was achieved for surface damage\cite{Paper:SB-pixel}, mostly because the related surfaces are up to some point accessible to the design tools of standard CMOS-processes. This does not hold for bulk damage. Therefore, strategies for improving the pixels in this respect are by far less obvious and the best solution might consist in reducing the operation temperature of the sensors despite of the non-trivial implications on the global concept of the vertex detectors. Doing the right technology choices requires a precise understanding on whether an observed increase of noise and leakage currents is caused by bulk damage or by surface damage. 

\section{Neutron irradiation and parasitic $\gamma$-rays}

In first order, one can obtain this understanding based on the above mentioned good choice of sources used for the sensor irradiation. By doing so, one neglects however the parasitic $\gamma$-radiation found in reactor neutron beams.  Fortunately, the related radiation doses are small: An upper limit of \mbox{$\lesssim \rm 100~krad$} per $\rm 10^{13}~n_{eq}/cm^2$ is typically communicated by the irradiation facilities. Elder MAPS designs, which tolerated $\lesssim 10^{13}~\rm n_{eq}/cm^2$, received ionizing doses of only $\lesssim \rm 100~krad$ during neutron irradiation. Those doses were accepted knowing that their presence caused a bias to our radiation tolerance measurements toward smaller values only. Still, the  effect of the ionizing doses was reduced by roughly one order of magnitude by not powering the sensors during the irradiation process\footnote{A substantial part of surface damage is caused by the generation of e/h-pairs in the $\rm SiO_2$. Electric fields evacuate the electrons from the $\rm SiO_2$ while the holes stay in place and create problems. By not powering the device, one removes the fields and eases the immediate recombination of the e/h-pair.}. 

\begin{figure}[t]
\begin{center}   
   
   \includegraphics[viewport= 2.2cm 0.5cm 30cm 19cm, clip, width=.98\columnwidth]{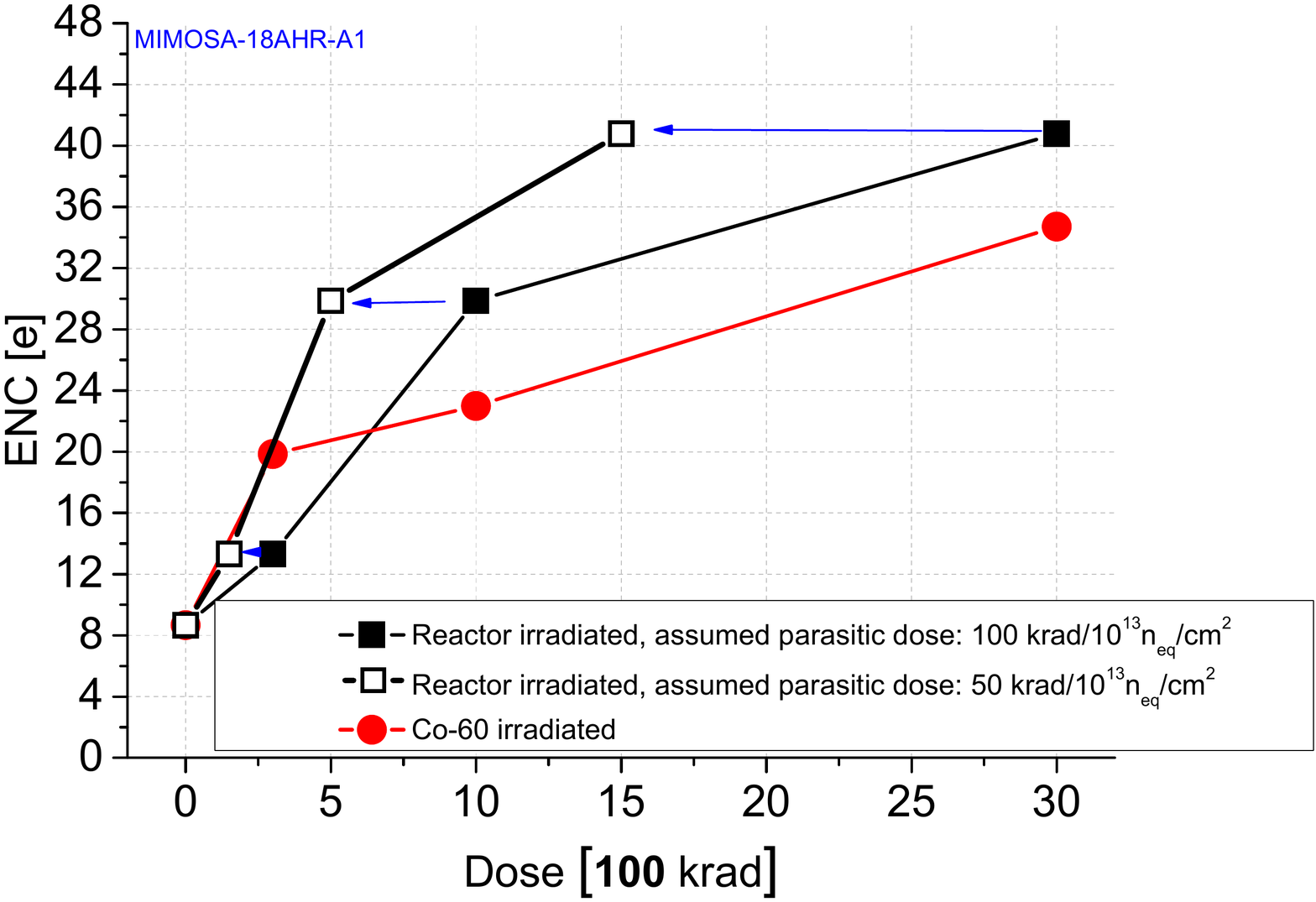}
   \caption{\label{Figure:Noise} Noise obtained after irradiating non-powered MIMOSA-18AHR with $\gamma$-rays. Measurements were taken with $\rm T_{sensor}=-3^{\circ}C$ and an integration time of $\sim 4~\rm ms$. The noise observed at the reactor-irradiated sensors is represented twice: Once, we assumed a parasitic $\gamma$-dose of $\rm 100~krad$ per $\rm 10^{13}~n_{eq}/cm^2$ and once, we assume $\rm 50~krad$ per $\rm 10^{13}~n_{eq}/cm^2$. The uncertainties are $<10\%$.}
	\end{center}
\end{figure}
The situation changed with the arrival of first sensors with partially depleted sensor, namely the prototype \mbox{MIMOSA-18AHR}. The latter resisted neutron fluencies of up to \mbox{$3 \cdot 10^{14}~\rm n_{eq}/cm^2$ \cite{Paper:1e14}.} During this irradiation in the neutron beam, the non-powered MIMOSA-18AHR was exposed to ionizing radiation doses of $\lesssim 3~ \rm Mrad$. If applied to a powered sensor, this dose exceeds the radiation tolerance of the device substantially. It was therefore not excluded that the increase of shot noise observed on the irradiated samples was caused by surface damage rather than by bulk damage. This holds in particular as, for the sake of simplicity, the pixel diodes of MIMOSA-18AHR are not equipped with guard rings deflecting possible leakage currents generated by surface damage. Given those facts, we decided to perform explicit test concerning a possible impact of the parasitic $\gamma$-rays on the device.

\section{Effect of parasitic $\gamma$-rays on the noise}

The tests were carried out with four MIMOSA-18AHR sensors provided by the PICSEL group of the IPHC Strasbourg. The sensor was manufactured in an $\rm AMS~ 0.35~\upmu m$ CMOS process. A $15~\rm \upmu m$ thick, high resistivity \mbox{($\rm \sim 1~k\Omega \cdot cm$)} epitaxial layer is used as active volume.  Three sensors were irradiated  with $\gamma$-rays from a $\rm ^{60}Co$-source at University Giessen. Doses of $300~\rm krad$, $1~\rm Mrad$, $3~\rm Mrad$ were applied and the sensors were intentionally not powered during irradiation in order to reproduce the conditions during neutron irradiation. The SB-pixels tested had a $10\rm \upmu m$ pitch and a $15\rm \upmu m^2$ collection diode.

The leakage current of those SB-pixels are too small for direct measurement. Moreover, they are not accessible via the pixel amplifiers as the related charge it is continuously removed by the forward biased diode. Our studies had therefore to rely on noise measurements. This was done knowing that most noise increase reflects additional shot noise created by radiation induced leakage currents. The noise was measured by taking 100 consecutive measurements of the dark signal of each pixel, performing correlated double sampling and computing the standard deviation of the data obtained. The average noise of all pixels are displayed in Figure \ref{Figure:Noise}, which also contains noise values observed in neutron irradiated chips. The anticipated $\gamma$-doses of the latter sensors cannot be indicated without additional assumptions. In the plot, we assumed once a pessimistic $\gamma$-dose of $\rm 100~krad$ per $\rm 10^{13}~n_{eq}/cm^2$ and once a dose of $\rm 50~krad$ per $\rm 10^{13}~n_{eq}/cm^2$. The ionizing doses indicated for the neutron irradiated sensors were obtained by multiplying the known neutron doses of the sensors ($\rm 3 \cdot 10^{13}~n_{eq}/cm^2$, $\rm 10^{14}~n_{eq}/cm^2$ and $\rm 3 \cdot 10^{14}~n_{eq}/cm^2$) with those factors and should not be confused with measured values.  

From the noise observed with the $\gamma$-irradiated sensors, one can conclude that this radiation causes substantial radiation damage also in unpowered MAPS. In case the true $\gamma$-dose follows the anticipated upper limit of $\rm 100~krad$ per $\rm 10^{13}~n_{eq}/cm^2$, the related surface damage may dominate the noise increase found in the neutron irradiated sensors. However, the noise observed in the $\gamma$-irradiated sensors overshoots the values observed in the neutron irradiated ones at some point. The most plausible explanation for this observation is that the true parasitic $\gamma$-dose of the neutron source amounts $ \rm 50~krad$ per $\rm 10^{13}~n_{eq}/cm^2$ or less. In any case, the impact of the parasitic radiation seems non-negligible unless a more precise dosimetry for the parasitic $\gamma$-doses proves that our assumptions on the parasitic doses are substantially too high. Until then, one concludes that the usual irradiation protocols do not provide reliable knowledge on the effect of bulk damage on the leakage current and the noise of MAPS.

\section{Suppressing the effect of parasitic $\gamma$-rays by annealing}
\begin{figure}[t]
\begin{center}   
   
 \includegraphics[viewport= 2cm 0cm 26.5cm 19cm, clip, width=.98\columnwidth]{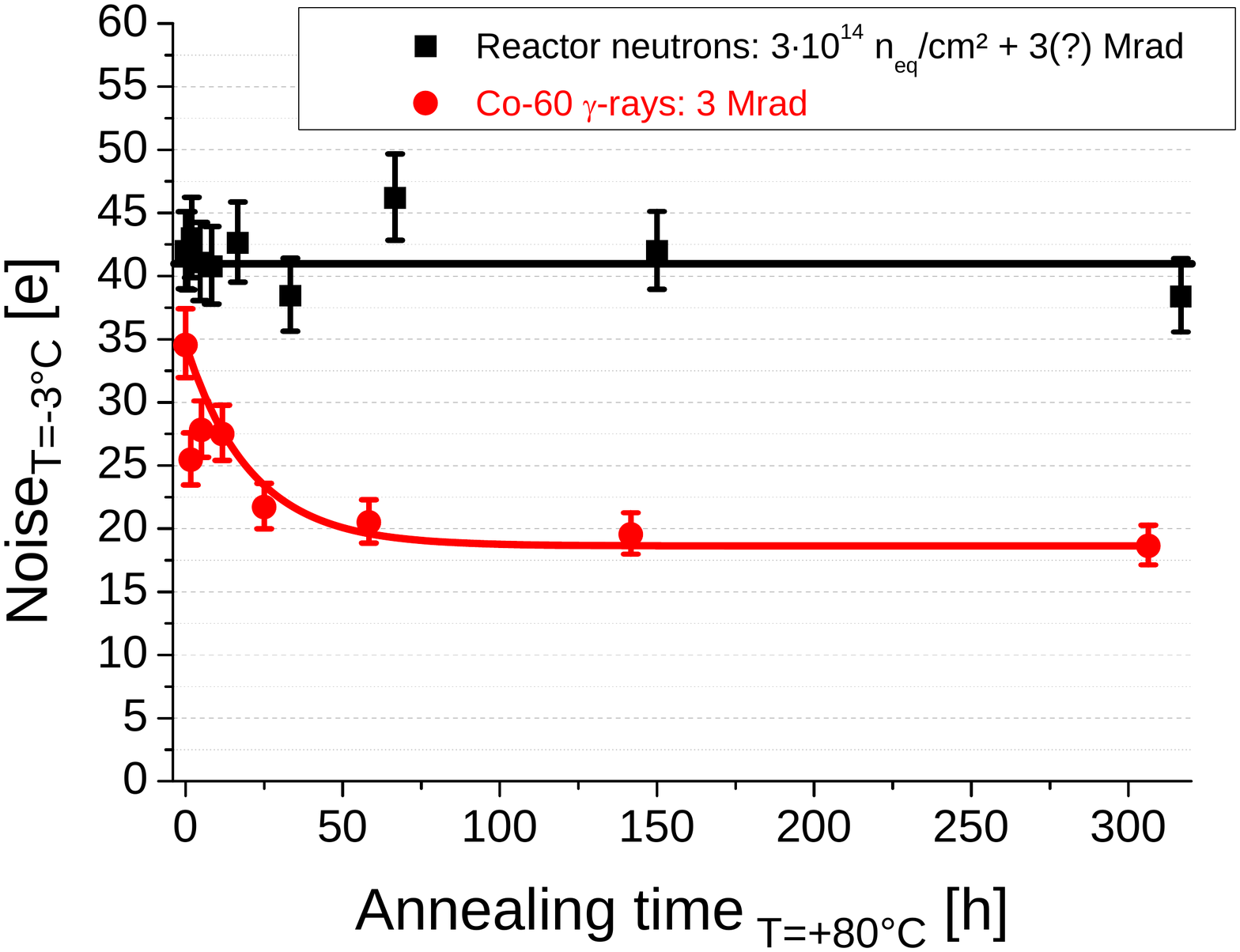}
   \caption{\label{Figure:Annealing} Noise measured during the annealing of a neutron- and a $\gamma$-irradiated sensor. See text.}
	\end{center}
\end{figure}
In a second step, we tried to suppress the contribution of the parasitic $\gamma$-rays by means of an additional forced thermal annealing. This accounts for the observation that noise generated by ionizing (X-ray) radiation damage is strongly reduced by an annaling at $T=80~\rm ^{\circ}C$ while the noise of neutron irradiated devices remains mostly stable \cite{Paper:Annealing}. In the plausible (but not obvious) case that the radiation damage caused by the $\gamma$-rays shows similar annealing behaviour, this procedure is expected to eliminate a substantial fraction of the surface damage while the bulk damage remains mostly unaffected. 

To test this hypothesis, we annealed a $\gamma$- and a neutron-irradiated sensor for 320 hours at $80~\rm ^{\circ}C$. As shown in Figure \ref{Figure:Annealing}, we observed the expected annealing behavior: The noise of the $\gamma$-irradiated device is reduced by about a factor of two. The noise of the neutron irradiated sensor remains mostly unchanged despite possible surface damage should be alleviated in analogy to what was observed in the $\gamma$-irradiated device. After subtracting the noise of the $\gamma$-irradiated device quadratically from the one of the neutron irradiated sensor, we conclude that the noise $N$ caused by isolated bulk damage from $3\cdot 10^{14}~\rm n_{eq}/cm^2$ should be dominant after annealing and amount $ 33~\rm e~ENC \lesssim$ $N$ $\rm \lesssim 42~e~ENC$. 

\section{Summary and conclusion}

Ionizing and non-ionizing radiation create different microscopic damages in MAPS, which manifest themselves in similar macroscopic effects (e.g. increased leakage current and shot noise). As alleviating the different kinds of damages requires different device optimization strategies, a precise knowledge on the origin of those effects is needed. One has therefore to exclude that the noise increase observed at neutron irradiated sensors is caused by the parasitic $\gamma$-doses of $\rm \lesssim 100~krad$ per $10^{13}~\rm n_{eq}/cm^2$ found in neutron beams.

To study the impact of parasitic $\gamma$-rays, we irradiated non-powered sensors with $\gamma$-doses as expected from neutron irradiations. In contrast to our initial assumptions, the parasitic $\gamma$-doses are found to create a noise increase, which is suited to explain the noise increase found at the neutron irradiated sensors. Consequently, additional tests steps are needed to estimate the impact of bulk damage. We find that a thermal annealing reduces the noise of $\gamma$-irradiated sensors while the noise in neutron irradiated sensors remains unchanged. From this, we conclude that the noise found in the latter sensors is indeed caused by bulk damage. We recommend to apply a similar annealing protocol whenever quantitative numbers on the consequences of non-ionizing radiation are required. 

Finally, we remind the reader that the noise measurements shown in this work were  carried out intentionally under unfavourable conditions ($4~\rm ms$ long integration time). Therefore, they don't reflect the performances of the by orders of magnitude faster devices foreseen for future vertex detectors. Moreover, as the $\gamma$-irradiations were carried out on non-powered sensors, no claim  on their regular tolerance to ionizing radiation is made.

%
			%
%
%

\section*{Acknowlegments}
This work was supported by HIC for FAIR and the BMBF(06FY9099I,05P12RFFC7).








\end{document}